# Time-resolved infrared spectroscopy with entangled photons and visible-light detection

Jasleen Kaur[1,2], Yu Mukai[1–3], Ryo Okamoto[1,2] & Shigeki Takeuchi[1,2]*

[1]Department of Electrical, Electronic, and Digital Science and Engineering, Kyoto University, Kyotodaigaku-katsura, Nishikyo-ku, Kyoto 615-8510, Japan.

[2]Photonic Quantum Sensing Science and Engineering Center, Graduate School of Engineering, Kyoto University, Kyoto, Japan

[3]Japan Science and Technology Agency, PRESTO, Gobancho, Chiyoda-ku, Tokyo 102-0076, Japan

*e-mail: takeuchi@kuee.kyoto-u.ac.jp

## Summary paragraph

Time-resolved infrared (TR-IR) spectroscopy reveals molecular structural changes and charge-carrier dynamics on their intrinsic timescales, in photosynthesis, catalysis and optoelectronic materials[1–4], but requires complex sources of ultrashort infrared pulses[5,6], cryogenically cooled detectors to overcome thermal background[7], and intense probes that can perturb the sample. Quantum infrared spectroscopy avoids these constraints: harnessing quantum interference between the generation processes of frequency-entangled photon pairs, it reads out the infrared response of a sample by detecting only the visible partner photon, for which efficient detectors exist[8–19]. To date, it has been limited to steady-state measurements. Here we demonstrate its ultrafast, time-resolved form with a picosecond visible pump pulse and a silicon detector. Each visible pulse generates photon pairs within a picosecond window; the infrared photon of a pair probes the excited sample at a set delay after excitation, and the transient response is read out from the count rate of its visible partner. We resolve picosecond carrier dynamics in a semiconductor saturable absorber mirror (SESAM) near 1.53 μm with a picowatt-level probe, unaffected by thermal background and by sample IR photoluminescence 50 times stronger than the probe. The approach is applicable to the mid-infrared[10–15], fingerprint region[16–19], where infrared detection is hardest, and to the terahertz region[20,21].

## Main

Time-resolved infrared (TR-IR) spectroscopy follows how the infrared absorption spectrum of a sample evolves in time. Because infrared absorption reports directly on free carriers and excitons in semiconductors[1] and on the making and breaking of chemical bonds[2,3], TR-IR captures the elementary events of photophysics, chemistry and biology on their intrinsic timescales[4]. Step-scan Fourier-transform spectroscopy reaches nanosecond time resolution[22]; the picosecond and femtosecond events that matter most require a pump–probe experiment with broadband ultrashort infrared pulses. Such an experiment is demanding on three fronts. The probe pulses must be produced by optical parametric amplification and difference-frequency generation driven by a high-power pulsed laser[5,6]. The infrared detectors must be cooled cryogenically to suppress thermal background noise and enable reliable signal detection[7]. And an adequate signal-to-noise ratio demands an intense infrared probe, which can itself perturb or damage the sample. Upconversion[23,24] and electro-optic sampling[25] shift the detection to silicon-based sensors but still require an infrared probe source and add an intense femtosecond gate pulse. Ultrafast TR-IR is thus one of the most informative probes of dynamics in matter—and one of the least accessible, confined to a small number of specialized laboratories.

Infrared spectroscopy itself, meanwhile, is being transformed by quantum optics. In a nonlinear quantum interferometer, a photon pair can be generated in either of two nonlinear crystals; when the two generation processes are indistinguishable, they interfere, and the response of a sample placed in the path of one photon is imprinted on the count rate of its partner[26,27]. This principle has enabled imaging[27–30], optical coherence tomography[31–33] and quantum infrared spectroscopy (QIRS)[8–17], in which an infrared photon probes the sample and only its visible partner is detected with a silicon detector. QIRS has been demonstrated in the mid-infrared from 2–5 μm[10–15], in the 6–12 μm fingerprint region[16,17], at terahertz frequencies[20,21], and has recently been extended to 20 μm[18,19]. It needs neither an infrared source nor an infrared detector, its probe is at the picowatt level, and, because only photons correlated with the detected partners contribute to the interference, uncorrelated infrared light such as thermal radiation or sample photoluminescence does not enter the signal. We note that, unlike spectroscopy based on entangled two-photon absorption[34,35], the sample here interacts with the probe linearly; the quantum resource serves only the read-out. Each of these features removes one of the obstacles listed above. However, QIRS has so far been limited to steady-state measurements.

The step from steady-state QIRS to a pump–probe measurement requires pulsed excitation of the photon-pair source. It was generally expected that the finite linewidth of a pulsed pump would degrade the spectral resolution of QIRS by weakening the frequency correlation within each pair[36,37]. We recently showed theoretically that the spectral resolution of quantum Fourier-transform infrared (QFTIR) spectroscopy[9] is set by the scan range of the interferometer rather than by the pump linewidth, and confirmed this experimentally with an apparatus in which continuous-wave and pulsed pumping could be compared under identical conditions[38]. That experiment, however, did not involve time-resolved measurements, which remained to be demonstrated.

Here we demonstrate time-resolved QIRS (TR-QIRS): ultrafast infrared spectroscopy with frequency-entangled photon pairs, in which only the visible photon is detected, with a silicon single-photon detector. Realizing it required overcoming a practical difficulty. The infrared probe generated by parametric down-conversion is extremely weak—of the order of $10^{-20}$ J per pulse —so the spatial and temporal overlap of pump and probe at the sample cannot be established by detecting the probe, as it is in conventional TR-IR. We developed instead an alignment procedure that uses the 1,064-nm fundamental, sent along the probe path, as a reference beam. Furthermore, we designed the optical set-up so that the delay between the excitation pulse and the infrared probe, and the path difference between the probe and its visible partner that produces the quantum interferogram, can be controlled precisely and independently (Methods). Using the TR-QIRS system thus developed, we resolved the carrier dynamics of a semiconductor saturable absorber mirror (SESAM) (Methods) at 1.53 µm— with a 7-ps instrument response and 25-cm$^{-1}$ spectral resolution, detecting only photons at 816 nm with a silicon single-photon detector. The probe fluence lies eight to nine orders of magnitude below the saturation fluence of the sample and far below the typical probe fluences used in conventional mid-IR pump-probe spectroscopy experiments[39,40]. The measurement also proved immune to the infrared photoluminescence of the excited sample, which is 50 times stronger than the probe, by virtue of its character as a form of quantum illumination[41]: only photons correlated with the detected visible photons contribute to the interference. The scheme carries over to the mid- and far-infrared with a suitable pump laser and nonlinear crystal[10–19].

## Principle of time-resolved quantum infrared spectroscopy

Figure 1a shows the scheme. An ultrashort pump pulse (SPDC-pump pulse) incident on the first nonlinear crystal generates a visible–infrared photon pair via spontaneous parametric down-conversion (SPDC). The visible–infrared photon pair is generated as a wave packet confined within the duration of the SPDC-pump pulse (Fig. 1b). The same pulse, after passing through the first crystal, can generate a pair in the second crystal with an equal probability. When these two generation processes are indistinguishable, they interfere: the rate at which pairs are produced rises or falls depending on the relative phase between the two processes, which is varied by changing the path length of the IR photon, $\Delta L$. Counting the visible photons with a silicon detector as $\Delta L$ is scanned therefore gives a fringe pattern, the quantum interferogram (Fig. 1c, Supplementary Section 1.1). Note that this is an interference between generation processes, not between light fields. If the infrared photon from the first crystal is lost at the sample—absorbed or not reflected into the second crystal—the two processes become partially distinguishable, and the fringe amplitude decreases. The Fourier transform of the quantum interferogram gives the spectrum of the infrared probe photons reflected from the sample; comparison with the spectrum obtained with a reference mirror then yields the reflectance spectrum, with a spectral resolution set by the scan range of $\Delta L$ (the QFTIR method; Fig. 1c)[9,38].

To follow the sample dynamics in time, a sample-excitation pulse is incident (Fig. 1a). It is generated from the SPDC-pump pulse by wavelength conversion, so the two pulses are intrinsically synchronized, and the probe photon arrives at the sample at a defined time. A delay line in the excitation arm sets the interval $\tau$ between the excitation pulse and the probe, so that the probe can be made to arrive after $(\tau > 0)$ or before $(\tau < 0)$ the excitation. The optical path difference $\Delta L$, which determines the relative phase between the two generation processes, is controlled independently of $\tau$. The quantum interferogram at each delay $\tau$ is recorded and a Fourier transform is performed to extract the time-resolved infrared reflectance spectrum of the sample (Fig. 1c, Supplementary Section 1.2).

The spectrally averaged reflectance can also be obtained by scanning $\Delta L$ over a short range around the centre burst of the interferogram (the short-scan method): the visibility of the fringes, compared with that measured with a reference mirror, gives the reflectance (Supplementary Section 1.3). Because far fewer points are needed, the acquisition is correspondingly faster. We use the short-scan method for the transient reflectance in Fig. 3 and the QFTIR method for the time-resolved spectra in Fig. 4.

## Experimental implementation

For this proof of principle, we have chosen the telecom band because a well-characterized sample with picosecond dynamics is available there. The experimental set-up is shown in Fig. 2 (Methods). Picosecond pulses at 1,064 nm (5.2 ps, 80 MHz) are split into two arms: one provides the sample-excitation pulse, the other is frequency doubled to 532 nm (4.4 ps) to provide the SPDC-pump pulse. The SPDC-pump pulse pumps a 0.5-mm lithium niobate crystal, which generates infrared photons at 1,529 nm and visible photons at 816 nm by type-I spontaneous parametric down-conversion (SPDC); the crystal is traversed twice in a Michelson-type nonlinear quantum interferometer (Fig. 2), which realizes the two-crystal scheme of Fig. 1a with a single crystal. The sample-excitation pulse is delayed by delay stage 1, which sets $\tau$, and is focused by the same lens as the probe onto the same spot on the SESAM sample. The excitation pulse with a peak fluence of around 120 μJ $cm^{-2}$, enough to induce saturable absorption (Methods), illuminates a spot of size 80 μm × 135 μm much larger than the 6-μm probe spot— so that the probe samples a uniformly excited region. $\Delta L$ is scanned with stage 2, and the visible photons are counted with a silicon single-photon counting module. In the low-gain regime the infrared probe power is of the order of picowatts, that is, about $10^{-20}$ J or a tenth of a photon per pulse, so its fluence at the sample is eight to nine orders of magnitude below the saturation fluence of the SESAM, which rules out probe-induced modulation. The reference-mirror interferogram has a visibility of about 40% (Supplementary Section 2, Fig. S1a).

## Transient reflectance of the saturable absorber mirror

We first demonstrate the ultrafast time-resolved observation of the reflectance change of the SESAM with the TR-QIRS set-up (Fig. 3). The delay $\tau$ is referenced to the temporal overlap of the excitation and probe pulses, determined from the fit in Fig. 3b (Supplementary Section 3). Figure 3a shows the quantum interferograms with the excitation on at $\tau = 2.6$ ps (after excitation) and at $\tau = -10.8$ ps (before excitation). Note that an offset from scattered excitation light has been subtracted from the counts (Supplementary Section 3.1, Fig. S2). The visibility of the fringes increased from 30.11 ± 0.09% to 32.62 ± 0.08%, corresponding to reflectance of 0.580 ± 0.004 and 0.681 ± 0.004 (an increase of 17.4 ± 0.9%), taking the visibility of 39.54 ± 0.08% measured with a reference mirror. This increase in reflectance is a signature of absorption bleaching in the saturable absorber layer.

Figure 3b shows the reflectance as a function of the delay $\tau$, scanned in steps of 0.67 ps, with the excitation on (green) and off (purple). With the excitation off, the reflectance stays constant within the noise. With the excitation on, by contrast, it changes markedly on a picosecond timescale: it rises around $\tau = 0$ and then relaxes to a level slightly above the pre-excitation value. We fitted the data (black line) with a double-exponential recovery, convolved with a Gaussian instrument response of full-width at half maximum $\Delta t_{\mathrm{IRF}}$ (Supplementary Section 3.2). The fit gives a fast recovery time $\tau_1 = 8.1 \pm 1.3$ ps, attributed to carrier capture at defect states, and a slow recovery time $\tau_2 = 77.9 \pm 8.3$ ps, attributed to recombination of remaining carriers. The fit also gives the instrument response, $\Delta t_{\mathrm{IRF}} = 7.0 \pm 0.4$ ps, which agrees with the quadrature sum, 6.8 ps, of the durations of the SPDC-pump pulse (4.4 ps) and the excitation pulse (5.2 ps). The baseline of the reflectance with the excitation on lies about 0.025 below that with the excitation off, as is evident at negative delays; we attribute this to laser heating of the SESAM by the excitation pulses (Supplementary Section 4.1). These results demonstrate that the TR-QIRS set-up resolves the ultrafast, picosecond-scale reflectance change associated with the carrier dynamics in the SESAM at 1.53 μm, detecting only visible photons at 816 nm with a silicon detector.

## Time-resolved infrared spectra

Beyond the spectrally averaged reflectance, TR-QIRS can follow the change of the reflectance spectrum itself by applying the QFTIR method at each delay. To this end, we extended the scan of $\Delta L$ to 0.4 mm, which gives a spectral resolution of 25 cm$^{-1}$ after Fourier transformation. Figure 4a shows the quantum interferograms with the excitation on and off at $\tau = 2.6$ ps, and Fig. 4b their Fourier transforms, which correspond to the spectra of the infrared probe photons reflected from the sample. The transient reflectance spectra in Fig. 4c are the squared ratio of the excitation-on to the excitation-off Fourier amplitudes at each delay, normalized to the same squared ratio at $\tau = -50.8$ ps; the interleaved excitation-off measurement cancels slow drifts of the set-up, and the normalization removes the delay-independent effect of laser heating (Supplementary Section 4.2).

The ratio of the normalized reflectance minus one gives the relative change in reflectance, $\Delta R(k,\tau)/R(k,\tau = -50.8\,\text{ps})$ as a function of wavenumber $k$ and delay $\tau$ (Fig. 4c). At negative delays, no change is observed. At $\tau = 2.6$ ps, the reflectance rises across the whole band, by 12% at 6,475 cm$^{-1}$ and by 19–22% between 6,500 and 6,600 cm$^{-1}$. Averaged over the band, the change is 19.4 ± 1.1%, in agreement within two standard deviations with the value of 17.4 ± 0.9% obtained with the short-scan method. As the delay increases, the change decreases across the whole band. At 15.9 ps it is already small at the low-wavenumber edge, where it was smallest from the outset, while it remains visible between 6,500 and 6,600 cm$^{-1}$. By 109 ps, the reflectance has almost returned to its initial value.

The data of Fig. 4c can also be read as the time response at each wavelength (Fig. 4d). The change in reflectance is larger at shorter wavelengths (higher wavenumbers), whereas the recovery time is similar at all wavelengths within the present signal-to-noise ratio, consistent with a recovery dominated by carrier capture at defect states, which does not depend on the probe wavelength[42]. These results demonstrate that TR-QIRS captures the ultrafast evolution of the infrared reflectance spectrum, again by detecting only visible photons with a silicon detector.

## Discussion

We now analyse the signal-to-noise ratio (SNR) of the present TR-QIRS system. Replacing the SESAM with a reference mirror, we recorded 200 interferograms and obtained 100 reflectance spectra from consecutive pairs; the SNR of the reflectance, defined as the mean divided by the standard error over these spectra, is 263 at 6,550 $cm^{-1}$ (Supplementary Section 5). This value is comparable to those reported for QFTIR with continuous wave pumping[15]. The SNR expected from the shot noise of the detected photons, 304, is close to the measured value, indicating that the present system is limited by photon shot noise rather than by technical noise.

A distinctive feature of TR-QIRS is its immunity to light that is uncorrelated with the detected visible photons: such light, whether thermal radiation or luminescence from the excited sample, does not contribute to the interference and therefore does not enter the signal. The SESAM provides a stringent test of this property, because under excitation it emits photoluminescence that overlaps the probe band. Figure 5 compares the photoluminescence spectrum, measured under the same excitation conditions as in Figs. 3 and 4, with the spectrum of the infrared probe: within the probe band (6,450–6,700 $cm^{-1}$), the photoluminescence is about 50 times stronger than the probe (Supplementary Section 6). Nevertheless, the transient reflectance of Figs. 3 and 4 was obtained without any correction for the photoluminescence. In a conventional

pump–probe measurement, a probe strong enough to be distinguished from this photoluminescence could itself affect the carrier dynamics and the luminescence of the SESAM. This immunity to background reflects the character of TR-QIRS as a form of quantum illumination[41]: by harnessing quantum entanglement and a quantum measurement, it distinguishes the probe photons from background light, so that only the photons correlated with the retained visible partner contribute to the signal. In TR-IR spectroscopy generally, the excitation pulse generates scattered light and the excited sample itself emits background at the probe wavelength, which are difficult to separate from the probe signal. TR-QIRS is therefore most powerful precisely in these situations, as well as in spectral regions where efficient infrared detectors are not available.

## Conclusion and outlook

We have demonstrated time-resolved quantum infrared spectroscopy: pump–probe infrared spectroscopy in which the probe is one photon (1.53 μm) of a frequency-entangled pair, generated with picosecond visible pulses and read out through its visible partner (816 nm) on a silicon detector, with neither an infrared source nor an infrared detector. The system achieves an instrument response of 7 ps and a spectral resolution of 25 $cm^{-1}$ with a probe of picowatt power. Using the short-scan method, we observed the transient reflectance of a semiconductor saturable absorber mirror and resolved a 17–19% rise followed by a two-component recovery with time constants of 8.1 ps and 78 ps. Using the QFTIR method at each delay, we further demonstrated the measurement of transient reflectance spectra over the 6,450–6,700 $cm^{-1}$ band at each delay, which agree with the spectrally averaged result of the short-scan method. The measurement was unaffected by infrared photoluminescence from the excited sample 50 times stronger than the probe, thanks to the property of TR-QIRS as a form of quantum illumination.

With an appropriate choice of SPDC-pump laser and nonlinear crystal, TR-QIRS can be extended to the mid-infrared from 2–5 μm[10–15], to the 6–20 μm fingerprint region[16–19] and to the terahertz region[20,21], where sources and detectors are a major bottleneck and the thermal background is significant[8]; steady-state QIRS has already been demonstrated in each of these regions with the same type of interferometer. The Michelson geometry used here measures the transient reflectance of the sample; a Mach–Zehnder-type nonlinear quantum interferometer, in which the probe traverses the sample once, gives the corresponding transient transmission measurement[8].

The time resolution is set by the pulse durations and can be extended to the femtosecond regime with a femtosecond visible laser. This would give access to intraband relaxation and coherent dynamics in semiconductors and to vibrational dynamics in molecules. There is no particular limit on the long-time side: the same scheme applies to nanosecond and microsecond dynamics with an optical or electronic delay.

Because the present system is shot-noise limited, its sensitivity and speed improve directly with the number of detected photons. A 100-fold increase in this number, through waveguide sources[43,44], operation at higher parametric gain[45,46] and full collection of the visible mode, would lower the detection limit to about 0.1% or shorten the measurement time 100-fold; for comparison, state-of-the-art classical TR-IR resolves absorbance changes of $10^{-4}$ or smaller[47]. The spectral window (here about 156 $cm^{-1}$ full-width at half maximum) can be widened with chirped quasi-phase-matched sources[11,43].

The scheme is not restricted to optical excitation. Any excitation that can be synchronized with the SPDC-pump pulse—electrical injection or magnetic pulses—can take the place of the sample-excitation pulse, since the probe needs only to arrive at a defined delay after the stimulus. Detecting the visible photons with a silicon-based sCMOS camera instead of a single-pixel detector would further enable time-resolved hyperspectral imaging[48], combining the spatial resolution of undetected-photon imaging with the temporal resolution demonstrated here.

Ultrafast infrared spectroscopy has required a high-power pulsed laser, an optical parametric amplifier chain and a cryogenically cooled detector. Here it is performed at room temperature with a compact picosecond visible laser, a bulk crystal and a silicon single-photon counter; no cryogenic cooling and no infrared spectrometer are needed, which makes the approach well suited to miniaturization. We anticipate that TR-QIRS will bring time-resolved infrared spectroscopy of photovoltaic and photocatalytic materials, reaction intermediates and biomolecules within reach of laboratories that do not specialize in ultrafast infrared optics.

# Methods

## SPDC-pump pulse

Pulses at 1,064 nm (5.2 ps, 80 MHz) from a mode-locked laser (aeroPulse, NKT Photonics) are split by a half-wave plate and polarizing beamsplitter into the excitation arm and the arm that generates the SPDC-pump pulse at 532 nm. In the latter arm, the 1,064-nm light is focused ($1/e^2$ focal spot diameter of about 84 μm) into a 3 × 3 × 15 mm lithium triborate (LBO) crystal (Edmund Optics) held at about 148 °C for non-critical type-I phase matching for the generation of 532 nm light. The average power at 532 nm is about 190 mW for an incident power of 1.3 W at 1,064 nm and the visible pulse duration is about 4.4 ps. Residual 1,064-nm light is removed with a 550-nm short-pass filter and a band-pass filter centred at 532 nm with a bandwidth of 4 nm (full-width at half maximum).

## Nonlinear quantum interferometer

The 532-nm pulses are focused into a 0.5-mm-thick lithium niobate crystal (CASTECH, cut angle 76.1°) tilted at 18.8° to the incident beam, such that IR photons (idler) centred at 1,529 nm and visible photons (signal) at 816 nm are generated via type-I SPDC. The nonlinear quantum interferometer is of Michelson type. Dichroic mirror DM2 (long-pass, edge 1,000 nm) transmits the IR and reflects the visible photons and the SPDC-pump pulse. The IR light is collimated by lens L4 and focused by lens L5 onto the sample or reference mirror and reflected. The visible photons and the SPDC-pump pulse are reflected by a concave mirror. The beams pass through the crystal a second time. Since the spatial mode of the IR photons is inverted due to focusing and reflection, while that of the visible photons is not, quantum interference only occurs for the collinearly emitted visible-IR photon pairs. Visible photons are transmitted by dichroic mirror DM1 (long-pass, edge 600 nm), collimated by lens L6, and the collinearly emitted photons participating in quantum interference are spatially filtered using an iris. After spectral filtering using a 600-nm long-pass filter and a band-pass filter (centre wavelength 850 nm and 40-nm full-width at half maximum at normal incidence, tilted to transmit at 816 nm), the visible photons are coupled into a fibre and detected by a silicon single-photon counting module (SPCM-AQRH-14-FC, Excelitas). The detected count rate was about $6 \times 10^4$ $s^{-1}$. The sample or reference mirror and lens L5 are mounted together on delay stage 2 (FS-1050UPX, Sigma Koki, step resolution 1 nm), which is scanned to vary $\Delta L$ while keeping the focus on the sample fixed.

## SESAM sample

The SESAM (SAM-1550-42-12ps-4.0-25.4g-c, BATOP optoelectronics) consists of a saturable absorber layer and a distributed Bragg reflector grown on a GaAs substrate designed for 1,550 nm. The specified saturation fluence is 40 μJ $cm^{-2}$, the relaxation time is 12 ps (manufacturer's value at 1,550 nm). The steady-state reflectance spectrum measured using QFTIR is given in Supplementary Section 2.

## Sample-excitation pulse and pump–probe delay

The sample-excitation pulse at 1,064 nm (average power about 0.4 W, pulse energy about 5 nJ, peak fluence 120 μJ cm$^{-2}$) passes through a retroreflector on delay stage 1 (OSMS20-35(Z), Sigma Koki; 200-μm round-trip step, 0.67-ps minimum delay increment) and is directed by a fixed pick-off mirror into lens L5, which focuses it onto the sample at the same position as the probe. The $1/e^2$ diameters of the excitation and probe beams are about 80 μm × 135 μm and 6 μm × 6 μm, respectively, so the probe samples a uniformly excited region. Because L5 and the sample move together on delay stage 2, the excitation and probe paths change equally when $\Delta L$ is scanned, and $\tau$ is independent of $\Delta L$. To set the spatial overlap, the fundamental 1,064 nm light was allowed to propagate along the probe path by switching off the LBO crystal temperature controller and removing short-pass filters. The foci of the two 1,064-nm light beams after L5 were superimposed using a beam profiler. To set the temporal overlap, a lithium niobate crystal was used in place of the sample: because the two 1,064-nm beams are non-collinear, a sum-frequency spot appears between the two second-harmonic spots only when the pulses overlap in time. The SESAM was mounted on a manual translation stage so that its surface coincided with the spatial and temporal overlap of excitation pulse and probe beam.

## Data acquisition and analysis

For each pump–probe delay, interferograms were recorded alternately with the excitation on and off, so that slow drifts of the interferometer and source affect both conditions equally. For the transient reflectance (Fig. 3) we recorded twenty excitation-on/off pairs of short-scan interferograms over 8 μm range in $\Delta L$, in steps of 100 nm (80 steps, 50 ms accumulation time per step). For the spectrally resolved measurement (Fig. 4) we recorded ten pairs of interferograms over 0.4 mm in $\Delta L$, in steps of 200 nm (2,000 steps, 50 ms accumulation time per step, measurement time of about 100 s each). With the excitation on, the mean count increases by about 960 counts per 50 ms accumulation time compared to excitation off (Supplementary Fig. S2). We attribute this to scattered excitation light; the increase in mean photon count is independent of $\Delta L$ and is therefore, not modulated by the interference. This offset was subtracted before the visibility was evaluated from the sinusoidal fit of the short-scan interferogram. For the Fourier-transform method no subtraction is needed, because a constant offset contributes only at zero wavenumber.

## Use of large language models

Claude (Anthropic; Fable 5.1 and Opus 5.5 models) was used to assist with writing the code for the curve fitting in Fig. 3b and analysing the spectral shift in Supplementary Fig. S3, restructuring the manuscript to the journal format and with language editing. The code and all suggestions were independently reviewed by the authors, who take full responsibility for the content.

## Data availability

Source data for Figs. 3–5 and Supplementary Figs. S1–S3 are provided with the paper. The raw interferograms are available from the corresponding author on reasonable request.

## Code availability

The analysis code is available from the corresponding author on reasonable request.

# Acknowledgements

This work was supported by MEXT Q-LEAP (JPMXS0118067634), JST ERATO Takeuchi Super Entanglement (JPMJER2402), JSPS KAKENHI (24H00195), JST PRESTO (JPMJPR2598) and a Grant-in-Aid for JSPS Fellows (25KJ1607). J.K. thanks H. Katsu for discussions.

# Author contributions

S.T. conceived the idea of time-resolved quantum infrared spectroscopy, led the project and acquired the funding. Y.M. and J.K. designed the experimental set-up. J.K. built the set-up, performed the measurements and analysed the data with the guidance of Y.M., R.O. and S.T. All authors discussed the results and their interpretation. J.K. wrote the first draft of the manuscript, and all authors revised it.

# Competing interests

The authors declare no competing interests.

# Additional information

Correspondence and requests for materials should be addressed to S.T.

# Figures

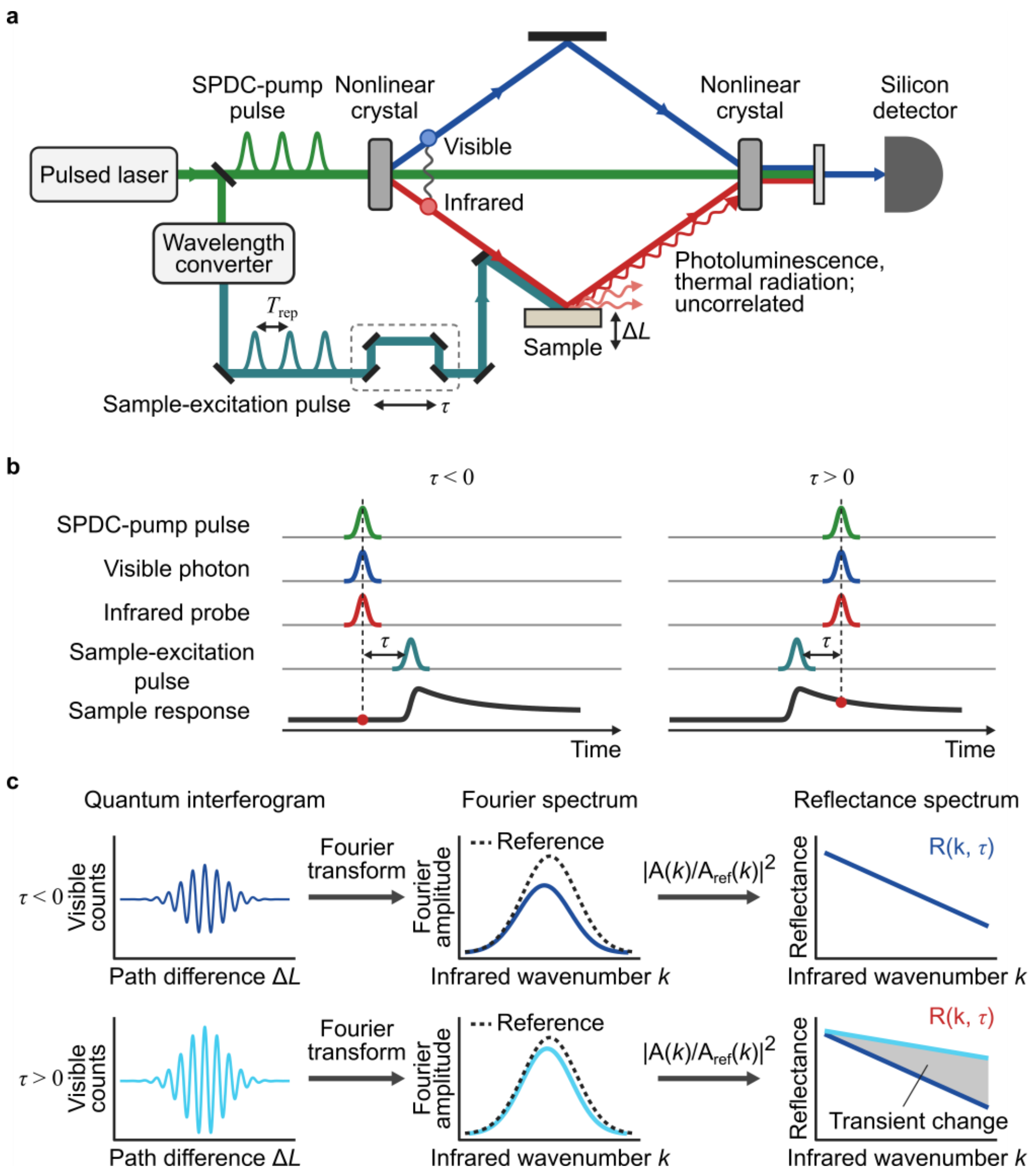

**Fig. 1 | Principle of time-resolved quantum infrared spectroscopy. a**, Scheme. A single pulsed laser provides the SPDC-pump pulse, which generates in a nonlinear crystal a frequency-entangled visible–infrared photon pair, localized in time within the pump pulse. The infrared photon (probe) is reflected from the sample and returned to a second crystal; its visible partner reaches the same crystal by a separate path. Because the pair can be generated with equal probability in either crystal, the two generation processes interfere, and the rate of visible photons, counted with a silicon detector, oscillates with the optical path difference $\Delta L$ that sets the relative phase between the two generation processes. This quantum interferogram is

converted into the infrared reflectance spectrum of the sample as shown in **c**. For time-resolved measurement, part of the SPDC-pump pulse is converted in wavelength into a sample-excitation pulse, which is incident at the same spot on the sample as the probe, a delay $\tau$ earlier, set by a delay line in the excitation arm. Recording the reflectance spectrum at each $\tau$ yields the time-resolved infrared spectrum. Photoluminescence and thermal radiation from the sample are uncorrelated with the detected visible photons and do not contribute to the interference. **b**, Timing, for a probe arriving before ($\tau < 0$ , left) and after ($\tau > 0$ , right) the excitation. The infrared probe and visible partner are generated only within the SPDC-pump pulse and therefore inherit their timing from it. The sample-excitation pulse, derived from the same laser, arrives $\tau$ earlier than the probe; $\tau$ is set in the excitation arm. Pulses repeat with the period $T_{\text{rep}}$. For $\tau < 0$ the probe arrives $|\tau|$ before the next excitation, that is, $T_{\text{rep}} - |\tau|$ after the previous one, and sees the relaxed sample; this is the same as excitation-off reference. For $\tau > 0$ the probe samples the response of the sample $\tau$ after excitation (black trace, red dot). **c**, From interferograms to time-resolved spectra. At each delay $\tau$ the quantum interferogram (left) is recorded against $\Delta L$ and Fourier transformed (middle). The ratio of its Fourier amplitude $A(k,\tau)$ to that measured with a reference mirror, $A_{\text{ref}}(k)$ (dashed), squared, gives the reflectance spectrum $R(k,\tau) = |A(k,\tau)/A_{\text{ref}}(k)|^2$ (right); the spectral resolution is set by the scan range $W$ of $\Delta L$, approximately $1/W$. The ratio of the spectra obtained with the probe arriving after and before excitation, $[R(k,\tau > 0)/R(k,\tau < 0)]$ gives the transient reflectance spectra. Curves are schematic.

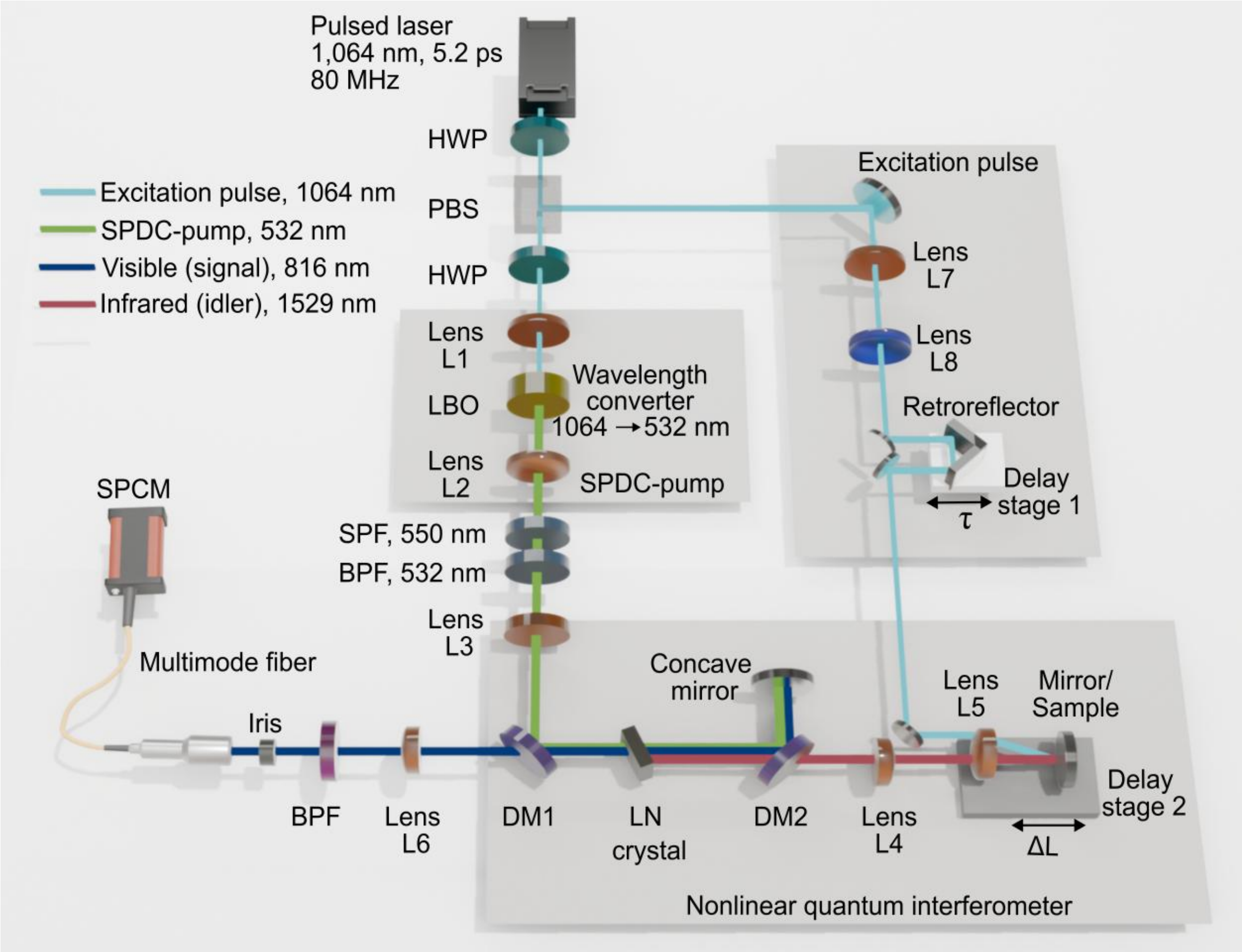

**Fig. 2 | Experimental set-up.** A 1,064-nm picosecond laser (5.2 ps, 80 MHz) is split by a HWP and PBS into the sample-excitation pulse and a part that is frequency-doubled in lithium triborate (LBO) crystal to produce the 532-nm SPDC-pump pulse. The SPDC pump drives a lithium niobate (LN) crystal in a Michelson-type nonlinear quantum interferometer generating visible (signal)-infrared (idler) photon pairs. Dichroic mirror DM2 transmits the 1,529-nm infrared photons, which are collimated by lens L4 and focused by lens L5 onto the sample or reference mirror placed on delay stage 2; and reflects the 816-nm visible photons and SPDC-pump light to a concave mirror. The sample-excitation pulse, delayed by $\tau$ with a retroreflector on delay stage 1, is focused by lens L5 onto the same spot as the infrared probe. The reflected visible photons, SPDC pump light and infrared photons pass through the crystal, where a second pair-generation process can occur; delay stage 2 scans the path difference $\Delta L$. Visible photons transmitted by DM1 are spatially and spectrally filtered (iris, band-pass filter with

centre wavelength of 850 nm, 40 nm full-width at half maximum, tilted to transmit 816 nm), coupled into a multimode fiber and detected by a silicon single-photon counting module (SPCM). HWP, half-wave plate; PBS, polarizing beamsplitter; SPF, short-pass filter; BPF, band-pass filter; DM, dichroic mirror; L1–L8, lenses.

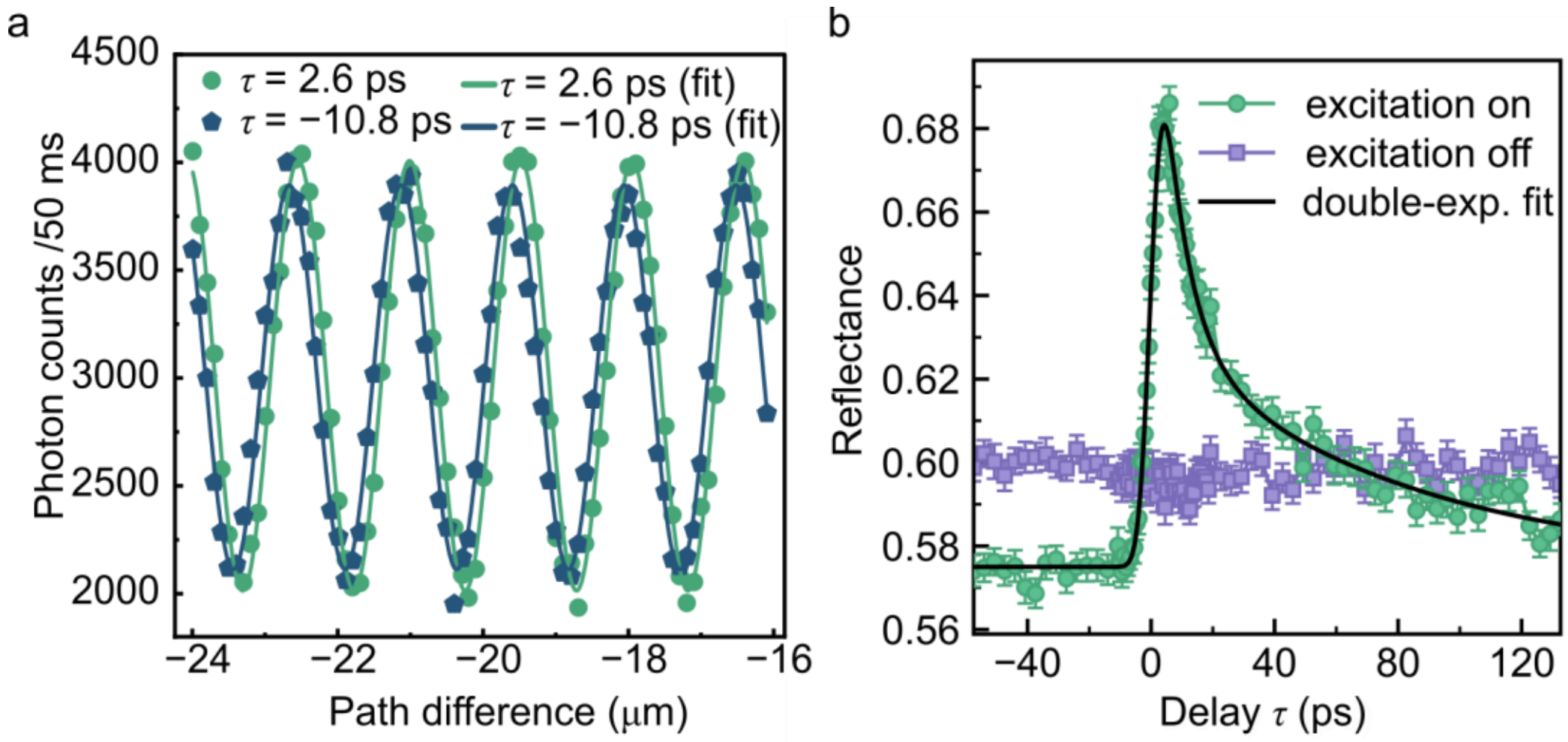

**Fig. 3 | Transient reflectance of the SESAM using the short-scan method. a**, Visible-photon interferograms (detected wavelength ~816 nm) with the excitation on at $\tau = 2.6$ ps (green) and $\tau = -10.8$ ps (blue) shown after subtraction of the offset from scattered excitation light (Methods, Supplementary Fig. S2). Lines are sinusoidal fits. The visibility increases from 30.11 ± 0.09% to 32.62 ± 0.08% (mean over 20 interferogram measurements) at $\tau = 2.6$ ps. **b**, Spectrally averaged reflectance versus delay with the excitation on (green) and off (purple), each the mean of 20 interferograms after offset correction (Methods). The black line is a fit to a double-exponential recovery, convolved with a Gaussian instrument response (Supplementary Section 3.2). It gives a fast recovery time of 8.1 ± 1.3 ps, a slower recovery component of 77.9 ± 8.3 ps and an instrument-response full-width at half maximum of 7.0 ± 0.4 ps. The lower excitation-on reflectance at negative delay is attributed to laser heating of the SESAM (Supplementary Section 4.1). Error bars, standard errors over twenty pairs of scans.

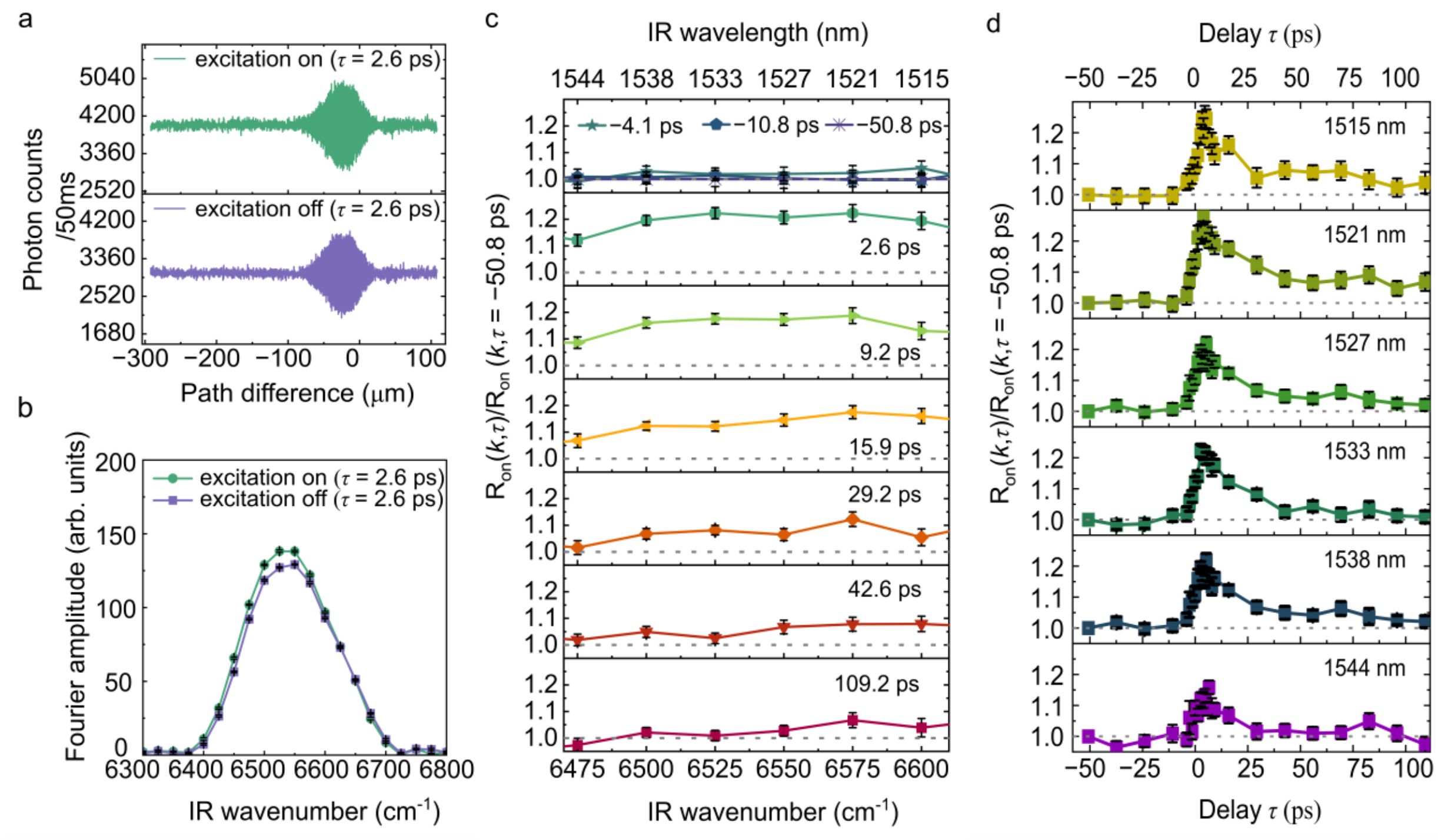

s

**Fig. 4 | Time-resolved infrared spectra of the SESAM. a**, Interferograms over a 0.4-mm path-difference scan with the excitation on (green) and off (purple) at $\tau = 2.6\,\mathrm{ps}$. **b**, Corresponding Fourier amplitudes (spectral resolution 25 cm$^{-1}$). **c**, Ratio of reflectance with excitation-on and excitation-off at delays from −51 ps to 109 ps, normalized to the ratio at $\tau = -50.8$ ps; this value minus one is the relative change of reflectance quoted in the text. **d**, The same data plotted against delay for six wavelengths between 1,515 and 1,544 nm. Error bars, standard errors over ten pairs of scans.

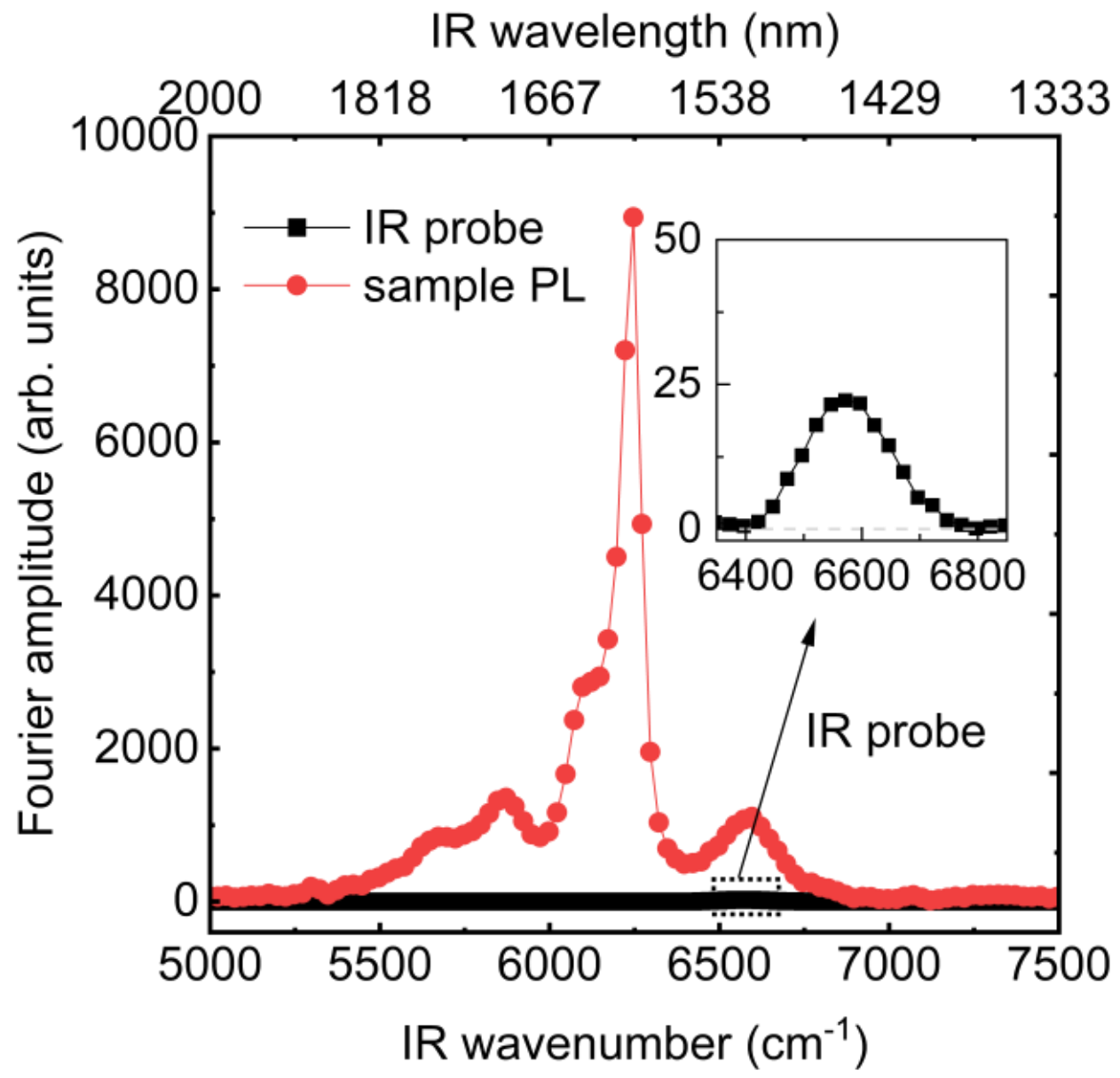

**Fig. 5 | Photoluminescence of the excited SESAM compared with the infrared probe.** Photoluminescence spectrum of the SESAM under excitation (red, peak fluence 120 μJ $cm^{-2}$) and spectrum of the infrared probe measured with the same set-up (black; inset, enlarged). The photoluminescence peaks at 1.6 μm with a shoulder near 1.52 μm; it exceeds the probe by more than two orders of magnitude at its peak and by about a factor of 50 within the probe band (6,450–6,700 $cm^{-1}$), yet does not contribute to the TR-QIRS signal because only visible photons are detected and only photons correlated with them contribute to the quantum interference.

**Supplementary Information**

# Time-resolved infrared spectroscopy with entangled photons and visible-light detection

Jasleen Kaur[1,2], Yu Mukai[1–3], Ryo Okamoto[1,2] & Shigeki Takeuchi[1,2]*

[1]Department of Electrical, Electronic, and Digital Science and Engineering, Kyoto University, Kyotodaigaku-katsura, Nishikyo-ku, Kyoto 615-8510, Japan.

[2]Photonic Quantum Sensing Science and Engineering Center, Graduate School of Engineering, Kyoto University, Kyoto, Japan.

[3]Japan Science and Technology Agency, PRESTO, Gobancho, Chiyoda-ku, Tokyo 102-0076, Japan

*e-mail: takeuchi@kuee.kyoto-u.ac.jp

**Contents**

# Supplementary Section 1. Theory of time-resolved quantum infrared spectroscopy

This section gives the general expression for the visible count rate as a function of path difference in a nonlinear quantum interferometer, the relation between the Fourier amplitude and the transient reflectance, and the extraction of reflectance using short-scan method.

## 1.1 Visible count rate

We consider frequency-entangled visible (signal) and infrared (idler) photon pairs generated in a nonlinear crystal by spontaneous parametric down-conversion (SPDC), driven by a pulsed pump (the SPDC pump) in the low-gain regime. In a Michelson-type nonlinear quantum interferometer, as in Fig. 2 of the main text, the SPDC pump passes through the crystal twice. Following refs S1 and S2, the visible count rate as a function of the optical path difference $\Delta L$, which sets the relative phase of the two generation processes, and at a fixed delay $\tau$ between the excitation pulse and the infrared probe, is given as,

$$P(\Delta L,\tau) \propto \iint dk_\mathrm{s}\, dk_\mathrm{i}\, |C\psi(k_\mathrm{s},k_\mathrm{i})\mu(k_\mathrm{s}+k_\mathrm{i})|^2 \left(2 + [r^*(k_\mathrm{i},\tau)]e^{-ik_\mathrm{i}\Delta L} + [r(k_\mathrm{i},\tau)]e^{ik_\mathrm{i}\Delta L}\right) \tag{S1}$$

where $k_\mathrm{s}$ and $k_\mathrm{i}$ are the wavenumbers in vacuum of the signal and idler, respectively, $\psi(k_\mathrm{s},k_\mathrm{i})$ is the phase-matching function, $\mu(k_\mathrm{s}+k_\mathrm{i}) = \mu(k_\mathrm{p})$ is the spectrum of the SPDC-pump pulse with $k_\mathrm{p}$ as wavenumber in vacuum, $C$ is a normalization constant, and $r(k_\mathrm{i},\tau)$ is the complex reflection coefficient of the sample seen by the IR probe at delay $\tau$. The first term in the braces gives the mean count rate, which depends on the joint spectral intensity of biphotons but not on the sample reflectance. The other two terms describe the quantum interference, whose visibility is proportional to $|r(k_\mathrm{i},\tau)|$, the modulus of the complex reflection coefficient of the sample at wavenumber $k_\mathrm{i}$ and delay $\tau$.

## 1.2 Quantum Fourier transform infrared spectroscopy method

With the Fourier transform defined as $A(k,\tau) = \left(\frac{1}{2\pi}\right) \int \mathrm{d}\Delta L \, P(\Delta L, \tau) \, e^{ik\Delta L}$, the Fourier amplitude of the quantum interferogram in equation (S1) at positive wavenumber $k$ is

$$A(k,\tau) = [r^*(k,\tau)] \int dk_{\mathrm{s}} |C\psi(k_{\mathrm{s}}, k)\mu(k_{\mathrm{s}} + k)|^2. \tag{S2}$$

Using the quantum Fourier transform infrared (QFTIR) spectroscopy method, the transient infrared reflectance at wavenumber $k$ and delay $\tau$ is given by,

$$R(k,\tau) = |A(k,\tau)/A_0(k)|^2 = |r^*(k,\tau)|^2. \tag{S3}$$

where $A_0(k)$ is the Fourier amplitude with reference-mirror measurement.

## 1.3 Short-scan method

In the short-scan method[S3], each interferogram is recorded over a short range of $\Delta L$ around the centre burst (8 μm in $\Delta L$, 80 steps of 100 nm, 50 ms accumulation time per step) and fitted with

$$P(\Delta L) = A \sin(k_{\mathrm{i}} \Delta L + \varphi) + y_0 \tag{S4}$$

where the fitting parameters are amplitude $A$, phase $\varphi$, mean counts $y_0$ and $k_{\mathrm{i}}$ is the centre idler wavenumber. The visibility $V = A/y_0$ obtained with the sample ($V_{\mathrm{SESAM}}$) and with the reference mirror ($V_{\mathrm{ref}}$) gives the spectrally averaged reflectance as $R(\tau) = \left[\frac{V_{\mathrm{SESAM}}(\tau)}{V_{\mathrm{ref}}}\right]^2$.

# Supplementary Section 2. Steady-state reflectance spectrum of sample

The steady-state reflectance spectrum of the SESAM is measured under the same experimental conditions using the QFTIR method with excitation off. We recorded ten interferograms each with the SESAM and with the reference mirror over 0.4 mm scan range in $\Delta L$, in steps of 200 nm, 50 ms accumulation time per step. Figure S1a shows the quantum interferograms recorded with the sample and with the reference mirror (visibilities about 30.3% and 39.7%, respectively), and Fig. S1b the corresponding Fourier spectra. The reflectance spectrum in Fig. S1c is the squared ratio of the Fourier amplitudes with the sample and with the reference mirror, both with the excitation off, at 25-cm$^{-1}$ resolution.

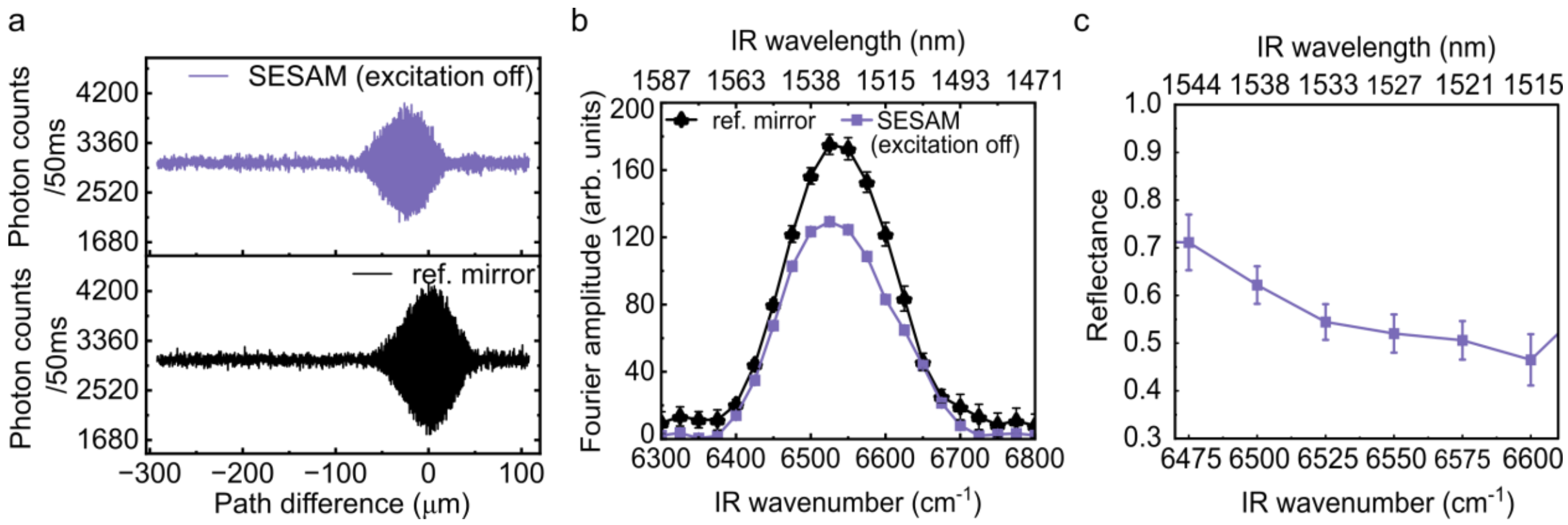

**Supplementary Fig. S1 | Steady-state reflectance spectrum of the SESAM.** a, Quantum interferograms with the reference mirror (black) and with the SESAM, excitation off (purple). b, Corresponding Fourier spectra. c, Steady-state reflectance spectrum with excitation off using QFTIR method (purple squares).

## Supplementary Section 3. Analysis of the transient reflectance measured with the short-scan method

This section describes the analysis of the transient reflectance with the excitation on, shown in Fig. 3 of the main text. We discuss the increase in the mean count with the excitation on, the double-exponential fitting model for the recovery of SESAM and the resulting parameters.

## 3.1 Mean count offset with excitation on

Fig. S2a shows the raw interferogram with excitation on and Fig. S2b with excitation off. The mean count $y_0$ increases by about 960 counts/50 ms (about 32%) with excitation on relative to the excitation-off value, independently of $\Delta L$ and $\tau$ without any change of the fringe amplitude. This is due to scattered excitation light reaching the detector. The offset, obtained from the difference between the excitation-on and excitation-off mean counts, was subtracted from the excitation-on interferograms before the visibility was evaluated; the interferograms in Fig. 3a are shown after this subtraction.

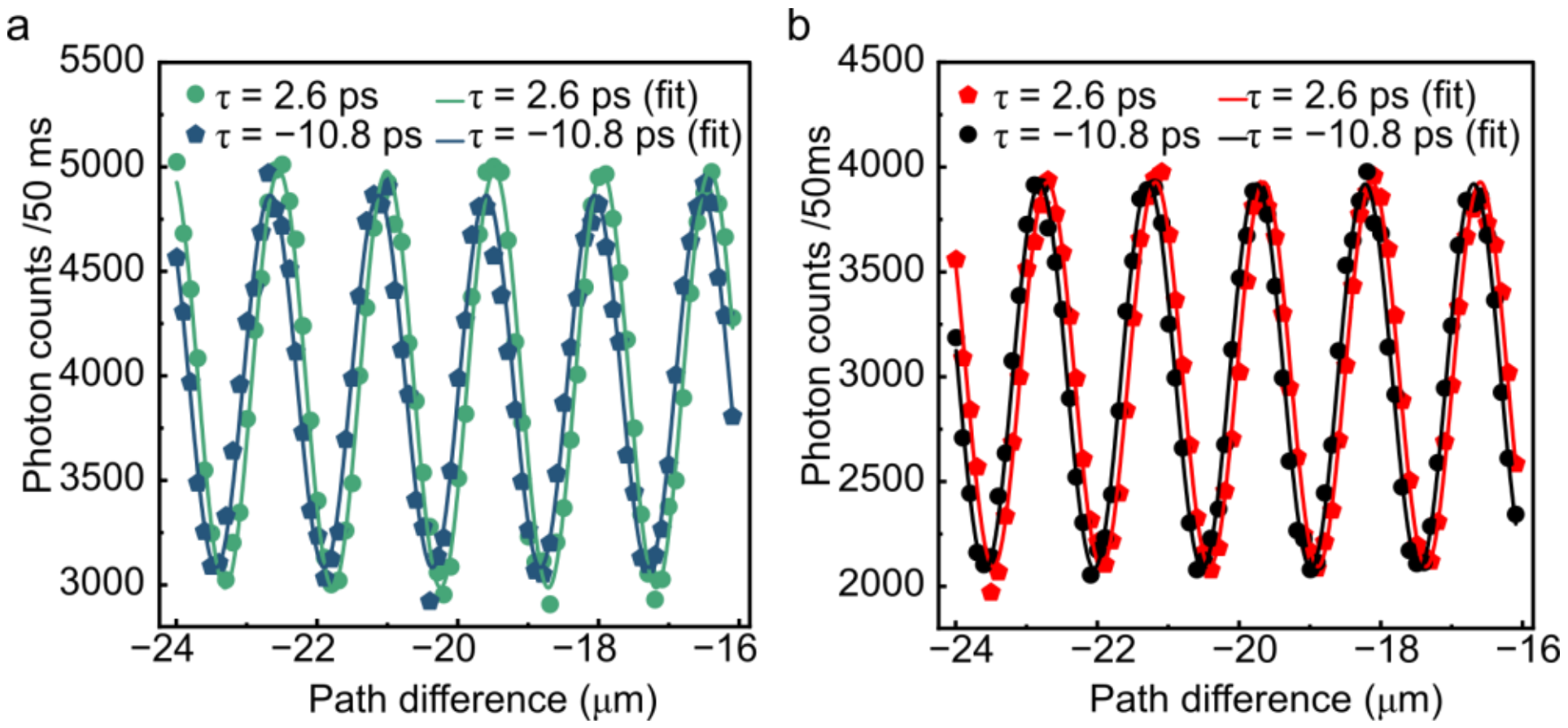

**Supplementary Fig. S2 | Mean count offset with excitation on and off.** Interferograms at $\tau = 2.6$ ps and $\tau = -10.8$ ps with the excitation on (a) and off (b), with sinusoidal fits. The mean count is higher with the excitation on which is attributed to additional uncorrelated light and is subtracted before evaluating the reflectance in Fig. 3.

## 3.2 Sample response function and fitting model

To quantify the transient response and the time resolution, we fitted the short-scan reflectance with the excitation on (Fig. 3b of the main text) with a bi-exponential response commonly used for SESAMs,

$$h(t) = \Theta(t)[\Delta R_1 \exp(-t/\tau_1) + \Delta R_2 \exp(-t/\tau_2)] \tag{S5}$$

where $\Theta$ is the Heaviside step function. Such a bitemporal recovery, with a fast component due to carrier capture at defect states and a slower one due to recombination of the remaining carriers, is characteristic of SESAMs[S4]. The measured trace is the convolution of $h(t)$ with the instrument response function (IRF) $G(t)$, which we take to be Gaussian,

$$G(t) = \frac{1}{\sigma\sqrt{2\pi}} \exp\left(-\frac{t^2}{2\sigma^2}\right), \quad \Delta t_{\mathrm{IRF}} = 2\sqrt{2\ln 2}\,\sigma \tag{S6}$$

where $\Delta t_{\mathrm{IRF}}$, the full width at half maximum (FWHM) of $G(t)$, defines the time resolution. The IRF convolved fit function is then

$$R(\tau) = R_0 + \Delta R_1 F_{\exp}(\tau - t_0; \tau_1) + \Delta R_2 F_{\exp}(\tau - t_0; \tau_2) \tag{S7}$$

with $F_{\exp}(t;\tau_i) = \frac{1}{2}\exp\left(\frac{\sigma^2}{2\tau_i{}^2} - \frac{t}{\tau_i}\right)\operatorname{erfc}\left(\frac{\sigma/\tau_i - t/\sigma}{\sqrt{2}}\right)$. Here, $\tau$ is the pump-probe delay, $R_0$ the reflectance before excitation and $t_0$ is time zero.

The parameters $(R_0, \Delta R_1, \tau_1, \Delta R_2, \tau_2, t_0, \Delta t_{\mathrm{IRF}})$ were obtained by a weighted nonlinear least-squares fit of equation (S7) to the mean reflectance at each delay, weighted by its standard error over 20 interferograms. Uncertainties are one standard deviation from the covariance matrix, and uncertainties of derived quantities were propagated including covariances.

The biexponential decay model gives $R_0 = 0.575 \pm 0.0008$, $\Delta R_1 = 0.106 \pm 0.008$, $\tau_1 = 8.1 \pm 1.3$ ps, $\Delta R_2 = 0.056 \pm 0.005$, $\tau_2 = 77.9 \pm 8.3$ ps, $t_0 = 0.00 \pm 0.17$ ps and $\Delta t_{IRF} = 7.0 \pm 0.4$ ps, with a reduced $\chi^2$ of 0.7. $R_0$ is consistent with the reflectance measured at $\tau = -10.8$ ps, $0.580 \pm 0.004$ (main text). The instrument response $\Delta t_{\mathrm{IRF}}$ agrees with the quadrature sum of 6.8 ps from the pulse durations of excitation pulse (5.2 ps) and SPDC pump pulse (4.4 ps) assuming Gaussian pulses. The deconvolved amplitude, $\Delta R_1 + \Delta R_2 = 0.162 \pm 0.010$, is the change that an infinitely short instrument response would record at time zero. The measured change in reflectance is lower, because the instrument response spreads it over about 7 ps, during which the fast component already partly recovers: the fitted curve gives a change of 0.100 between $\tau = -10.8$ ps and $\tau = 2.6$ ps, consistent with the measured 0.101 ± 0.006 (from the values in the main text).

## Supplementary Section 4. Thermal effects of excitation pulse on sample

This section discusses the origin of the offset of reflectance at negative delays between the excitation-on and off (Fig. 3b of the main text) and the normalization used in Figs. 4c,d.

### 4.1 Offset in transient reflectance at negative delay

At negative delays, the reflectance with excitation on lies about 0.025 below the excitation-off value (Fig. 3b). Also, the comparison of Fourier spectra with the excitation on and off shows that the spectrum with the excitation-on is shifted by 5.5 ± 0.1 $cm^{-1}$ to lower wavenumber, independently of delay (Fig. S3). This is attributed to laser heating of the sample by excitation pulses at 80-MHz repetition rate that modifies the resonance of the Bragg reflector in SESAM[S5]. The reflectance with excitation-on obtained by shifting the steady-state reflectance spectrum shown in Fig. S1c, by 5.5 $cm^{-1}$ is lower by 0.010 ± 0.004 compared to the excitation off value (weighted average over the probe band). This is of the same order as the observed difference of 0.025 which supports the origin of offset to be thermal.

### 4.2 Normalization of the transient spectra in QFTIR

At each delay, interferograms with the excitation on and off were recorded in alternation, and the Fourier transform amplitude of ten such measurements were averaged (Methods). In Fig. 4c (main text), the squared ratio of the excitation-on and excitation-off Fourier amplitudes is normalized to its value at $\tau = -50.8$ ps that can be expressed as

$$\frac{\left[\frac{A_{\mathrm{on}}(k,\tau)}{A_{\mathrm{off}}(k,\tau)}\right]^2}{\left[\frac{A_{\mathrm{on}}(k,\tau=-50.8\ \mathrm{ps})}{A_{\mathrm{off}}(k,\tau=\ -50.8\ \mathrm{ps})}\right]^2} = \left[\frac{A_{\mathrm{on}}(k,\tau)}{A_{\mathrm{on}}(k,\tau=-50.8\ \mathrm{ps})}\right]^2 \times \left[\frac{A_{\mathrm{off}}(k,\tau=-50.8\ \mathrm{ps})}{A_{\mathrm{off}}(k,\tau)}\right]^2 \quad \text{(S8)}$$

where $A_{\mathrm{on(off)}}(k,\tau)$ is the Fourier amplitude with excitation-on (off) at wavenumber $k$ and delay $\tau$. The second factor, a ratio of two excitation-off spectra, corrects slow drifts of the set-up between the measurements at the two delays. In the first factor, the heating-induced change of the excitation-on spectrum, which is independent of delay (Fig. S3), cancels, leaving the transient change caused by the excitation.

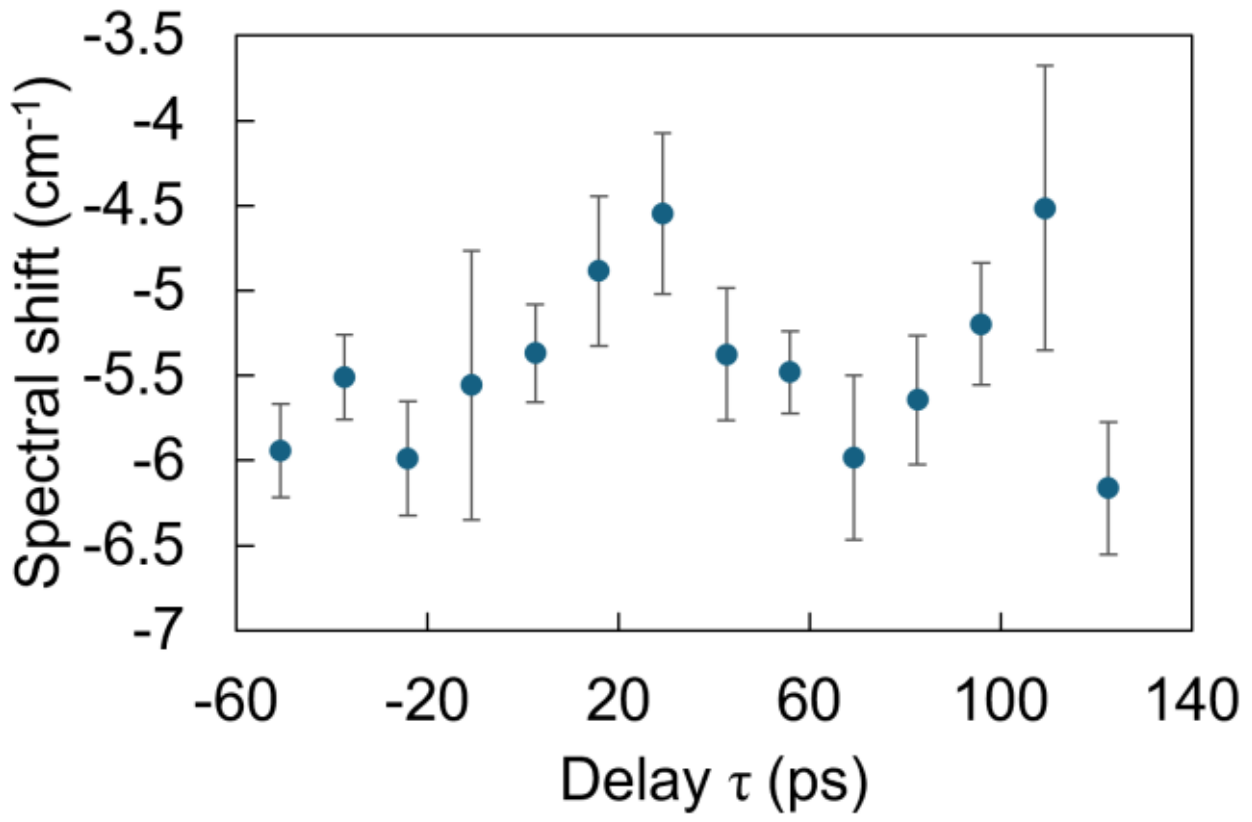

**Supplementary Fig. S3 | Heating-induced spectral shift.** Shift of the Fourier spectrum with the excitation on relative to the excitation-off spectrum, obtained at each delay by fitting each excitation-on spectrum as a shifted and scaled copy of the excitation-off spectrum recorded immediately before it (mean ± standard error over ten pairs). The shift is independent of delay, with a weighted mean of −5.5 ± 0.1 cm$^{-1}$ (to lower wavenumber).

## Supplementary Section 5. Sensitivity of QFTIR measurement

This section estimates the signal-to-noise ratio (SNR) of the reflectance obtained from the QFTIR measurement in our setup and compares it with the shot-noise limit. The SNR is defined as $\mu/\delta$, where μ is the mean and $\delta$ the standard error of the mean of reflectance values measured with the reference mirror in place of the sample. It was evaluated from a separate set of 200 reference-mirror interferograms (0.4 mm scans in $\Delta L$ with 200 nm step size, 50 ms accumulation time per step), recorded independently of the data in the main text; the visibility in this set was 0.337. The squared ratio of the Fourier amplitudes of each pair of consecutive interferograms gives one reflectance value, ideally equal to 1, and the resulting 100 values give μ and $\delta$. We obtain SNR ≈ 263 at 6,550 $cm^{-1}$. The sensitivity, minimum detectable change in reflectance, defined as $3\delta/\mu$ = 3/SNR is estimated as 1.14% at 25-$cm^{-1}$ resolution.

If photon shot noise is the only noise source, the SNR of the Fourier amplitude of a single interferogram in one spectral element near the centre of the SPDC spectrum is limited to

$$\mathrm{SNR_{th}} = \eta \sqrt{\frac{N_{\mathrm{total}}}{2} \frac{\Delta\tilde{\nu}}{\gamma}} \tag{S9}$$

where $\eta$ is the interference visibility, $N_{\mathrm{total}}$ the number of photons detected in one interferogram, $\Delta\tilde{\nu}$ the spectral resolution and γ the full-width at half maximum spectral width of the down-converted light. The ratio $\Delta\tilde{\nu}/\gamma$ is the inverse of the number of spectral elements over which the signal is distributed Whereas the shot noise in each spectral element scales as the square root of the total photon number. The additional factor $1/\sqrt{2}$ arises because shot noise is distributed equally over the two quadratures of the complex Fourier coefficient, and only the noise component parallel to the signal changes the measured amplitude; the perpendicular component changes only its phase. The measured SNR was evaluated at the band centre, 6,550 $cm^{-1}$, where the signal per spectral element is set by the effective spectral width, about 1.06 times the FWHM for a Gaussian spectrum.

For our QFTIR measurement, with $N_{\mathrm{total}}$ = 2,000 steps × 2,850 counts = $5.7 \times 10^6$, $\Delta\tilde{\nu}$ = 25 $cm^{-1}$ and γ ≈ 1.06 × 156 $cm^{-1}$, gives an SNR of about 86 for the Fourier amplitude of one interferogram. Each reflectance value is the squared ratio of two such amplitudes, which multiplies the relative noise by $2\sqrt{2}$, and averaging 100 values gives the shot-noise-limited

SNR of the reflectance as $86 \times \sqrt{100}/(2\sqrt{2}) \approx 304$. The measured value of 263 is within a factor of about 1.16 of this estimate. This indicates that the present system is limited by photon shot noise rather than by technical noise. A 100-fold increase in the number of photons would lower the detection limit to about 0.1%.

## Supplementary Section 6. Photoluminescence measurement scheme

The infrared photoluminescence of the sample (Fig. 5, main text) was measured with the same set-up, with the SPDC pump blocked and the excitation on (average power about 0.4 W, pulse energy about 5 nJ, peak fluence 120 μJ $cm^{-2}$). Photoluminescence emitted along the probe path, transmitted by dichroic mirrors DM2 and DM1 and collimated by lens L6, was coupled into a polarization-maintaining fibre and analysed with a separately constructed Michelson interferometer consisting of a fixed mirror and a mirror on a motorized stage (FS120-50UPX, Sigma Koki; step resolution 1 nm). After residual excitation light was removed with 1,450-nm and 600-nm long-pass filters, the output was coupled into a single-mode fibre and detected with a superconducting nanowire single-photon detector (SSPD, Scontel/Tokyo Instruments; bias current 19.5 μA), and the counts were recorded with a photon counter (SR400, Stanford Research Systems). Interferograms were recorded over a path-difference range of 0.4 mm with a step size of 200 nm (100 ms integration time per step), giving a spectral resolution of 25 $cm^{-1}$, as in the QFTIR measurements. The interferogram recorded with the excitation on was Fourier transformed to obtain the photoluminescence spectrum. For comparison, the probe spectrum was measured in the same way with the excitation blocked and the SPDC pump on. The photoluminescence peaks at 1.6 μm, where it could be removed by spectral filtering, but its shoulder overlaps the probe band (6,450–6,700 $cm^{-1}$), where it is about 50 times stronger than the probe; at its peak it is more than two orders of magnitude stronger.

# Supplementary References